\documentstyle[12pt,epsfig]{article}

\catcode`\@=11
\long\def\@makefntext#1{
\protect\noindent \hbox to 3.2pt {\hskip-.9pt  
$^{{\ninerm\@thefnmark}}$\hfil}#1\hfill}		%CAN BE USED 

\def\@makefnmark{\hbox to 0pt{$^{\@thefnmark}$\hss}}  %ORIGINAL 
	
\def\ps@myheadings{\let\@mkboth\@gobbletwo
\def\@oddhead{\hbox{}
\rightmark\hfil\ninerm\thepage}   
\def\@oddfoot{}\def\@evenhead{\ninerm\thepage\hfil
\leftmark\hbox{}}\def\@evenfoot{}
\def\sectionmark##1{}\def\subsectionmark##1{}}

\renewcommand{\thefootnote}{\fnsymbol{footnote}}

\newcounter{sectionc}\newcounter{subsectionc}\newcounter{subsubsectionc}
\renewcommand{\section}[1] {\vspace*{0.6cm}\addtocounter{sectionc}{1} 
\setcounter{subsectionc}{0}\setcounter{subsubsectionc}{0}\noindent 
	{\normalsize\bf\thesectionc. #1}\par\vspace*{0.4cm}}
\renewcommand{\subsection}[1] {\vspace*{0.6cm}\addtocounter{subsectionc}{1} 
	\setcounter{subsubsectionc}{0}\noindent 
	{\normalsize\it\thesectionc.\thesubsectionc. #1}\par\vspace*{0.4cm}}
\renewcommand{\subsubsection}[1]
{\vspace*{0.6cm}\addtocounter{subsubsectionc}{1}
	\noindent {\normalsize\rm\thesectionc.\thesubsectionc.\thesubsubsectionc. 
	#1}\par\vspace*{0.4cm}}

\newcounter{appendixc}
\newcounter{subappendixc}[appendixc]
\newcounter{subsubappendixc}[subappendixc]

\renewcommand{\appendix}[1] {\vspace*{0.6cm}
        \refstepcounter{appendixc}
        \setcounter{figure}{0}
        \setcounter{table}{0}
        \setcounter{equation}{0}
        \renewcommand{\thefigure}{\Alph{appendixc}.\arabic{figure}}
        \renewcommand{\thetable}{\Alph{appendixc}.\arabic{table}}
        \renewcommand{\theappendixc}{\Alph{appendixc}}
        \renewcommand{\theequation}{\Alph{appendixc}.\arabic{equation}}
        \noindent{\bf Appendix \theappendixc #1}\par\vspace*{0.4cm}}

\def\abstracts#1{{
	\centering{\begin{minipage}{12.2truecm}\footnotesize\baselineskip=12pt\noindent
	\centerline{\footnotesize ABSTRACT}\vspace*{0.3cm}
	\parindent=0pt #1
	\end{minipage}}\par}} 

\renewenvironment{thebibliography}[1]
	{\begin{list}{\arabic{enumi}.}
	{\usecounter{enumi}\setlength{\parsep}{0pt}
\setlength{\leftmargin 1.25cm}{\rightmargin 0pt}
	 \setlength{\itemsep}{0pt} \settowidth
	{\labelwidth}{#1.}\sloppy}}{\end{list}}

\newcounter{itemlistc}
\newcounter{romanlistc}
\newcounter{alphlistc}
\newcounter{arabiclistc}

\newcommand{\fcaption}[1]{
        \refstepcounter{figure}
        \setbox\@tempboxa = \hbox{\footnotesize Fig.~\thefigure. #1}
        \ifdim \wd\@tempboxa > 6in
           {\begin{center}
        \parbox{6in}{\footnotesize\baselineskip=12pt Fig.~\thefigure. #1}
            \end{center}}
        \else
             {\begin{center}
             {\footnotesize Fig.~\thefigure. #1}
              \end{center}}
        \fi}

\newcommand{\tcaption}[1]{
        \refstepcounter{table}
        \setbox\@tempboxa = \hbox{\footnotesize Table~\thetable. #1}
        \ifdim \wd\@tempboxa > 6in
           {\begin{center}
        \parbox{6in}{\footnotesize\baselineskip=12pt Table~\thetable. #1}
            \end{center}}
        \else
             {\begin{center}
             {\footnotesize Table~\thetable. #1}
              \end{center}}
        \fi}

\def\@citex[#1]#2{\if@filesw\immediate\write\@auxout
	{\string\citation{#2}}\fi
\def\@citea{}\@cite{\@for\@citeb:=#2\do
	{\@citea\def\@citea{,}\@ifundefined
	{b@\@citeb}{{\bf ?}\@warning
	{Citation `\@citeb' on page \thepage \space undefined}}
	{\csname b@\@citeb\endcsname}}}{#1}}

\newif\if@cghi
\def\cite{\@cghitrue\@ifnextchar [{\@tempswatrue
	\@citex}{\@tempswafalse\@citex[]}}
\def\citelow{\@cghifalse\@ifnextchar [{\@tempswatrue
	\@citex}{\@tempswafalse\@citex[]}}
\def\@cite#1#2{{$\null^{#1}$\if@tempswa\typeout
	{IJCGA warning: optional citation argument 
	ignored: `#2'} \fi}}

 1
 1
 1

\font\ninerm=cmr9

\def\beq{\begin{equation}}

\def\eeq{\end{equation}}

\def\as{\alpha_s}

\def\bea{\begin{eqnarray}}

\def\eea{\end{eqnarray}}

\begin{document}
\begin{flushright}
DESY 97-156\\
hep-ph/9708282
\end{flushright}
\vspace{.5truecm}
\centerline{\normalsize\bf ON HEAVY QUARKS PHOTOPRODUCTION}
\baselineskip=16pt
\centerline{\normalsize\bf AND \mbox{\boldmath $c\to D^*$} FRAGMENTATION
FUNCTIONS$^\clubsuit$}
\footnotetext{$^\clubsuit$Talk given at the Ringberg Workshop ``New Trends in
HERA Physics'', Ringberg Castle, Tegernsee, Germany, 25--30 May 1997.}
%\baselineskip=16pt
%\centerline{\normalsize\bf MANUSCRIPT BY COMPUTER}
%\centerline{\footnotesize\sf (For subsequent 20\% photoreduction
%to 17.8 $\times$ 11.9 cm text area)\footnote{The \LaTeX\ source
%file for this document may be used as a template for your
%article, and can be requested by e-mailing {\sf
%worldscp@singnet.com.sg}.}}

\vspace{.6truecm}
%\vspace*{0.6cm}
\centerline{\footnotesize MATTEO CACCIARI}
\baselineskip=13pt
\centerline{\footnotesize\it Deutsches Elektronen-Synchrotron DESY}
\baselineskip=12pt
\centerline{\footnotesize\it D-22603, Hamburg, Germany}
\centerline{\footnotesize E-mail: Matteo.Cacciari@desy.de}

\vspace{.9truecm}
%\vfill
%\vspace*{0.9cm}
\abstracts{The state of the art of the theoretical calculations for
heavy quarks photoproduction is reviewed. The full fixed order next-to-leading 
order massive calculation and the resummation of large $\log(p_T/m)$ terms 
for differential cross sections 
are described. The implementation of a non-perturbative fragmentation function
describing the $c\to D^*$ meson transition is also discussed.}
 
%\vspace*{0.6cm}
\normalsize\baselineskip=15pt
\setcounter{footnote}{0}
\renewcommand{\thefootnote}{\alph{footnote}}
\section{Introduction}
Heavy quarks production processes provide a powerful insight into our
understanding of Quantum Chromodinamics. The large mass of the heavy quark can
make the perturbative calculations reliable, even for total cross sections, by
cutting off infrared singularities and by 
setting a large scale at which the strong coupling can be
evaluated and found -- possibly -- small enough. On the experimental side, the 
possibility to
tag heavy flavoured hadrons by means of microvertex detectors can on the other
hand provide accurate measurements.

All these potentialities must of course be matched by accurate enough 
theoretical evaluations of the production cross section. In this talk I shall
describe the state of the art of such calculations for heavy quarks 
photoproduction. I shall first review the next-to-leading order (NLO) QCD
evaluations recently presented by Frixione, Mangano, Nason and Ridolfi.
These calculations, available for total cross
sections, one-particle  and two-particles distributions, are now a consolidated
result and provide a benchmark for future developments.

Large logarithms  appear in the  NLO fixed order calculations and
potentially make it less reliable in some regimes: $\log(S/m^2)$ 
and $\log(p_T^2/m^2)$ become large when the center of mass energy $\sqrt{S}$ or
the transverse  momentum $p_T$ of the
observed quark is much larger than its mass. I shall describe the resummation of
$\log(p_T^2/m^2)$ terms, leaving the high energy resummation to Marcello Ciafaloni's
talk\cite{ciafaloni}.

The perturbative fragmentation function
technique used in the resummation of the large $\log(p_T^2/m^2)$ terms has a
non-perturbative extension which can be used to describe the
transition from $c$ quarks to $D^*$ mesons. I shall therefore also discuss 
the determination
of these non-perturbative fragmentation functions and their inclusion into the
heavy quarks photoproduction calculation, showing a comparison with data from
HERA.

\section{Fixed Order NLO Calculation}
Heavy quarks photoproduction at leading order in the strong coupling $\as$
looks a very simple process: only the tree level diagram $\gamma g \to Q\bar Q$
contributes at the partonic level, and the final answer for the total cross
section 
%(to be convoluted with the gluon distribution function) reads
%\beq
%$
%\hat\sigma_{\gamma g}(\rho) = 2\pi\alpha\as e_Q^2 \rho\beta\left[\left(
%1+\rho-{\rho^2\over 2}\right){1\over\beta}\ln{{1+\beta}\over{1-\beta}}
%-1-\rho\right]
%%\eeq
%$
%with $\rho=4m^2/S$ and $\beta=\sqrt{1-\rho}$, $m$ being the heavy quark mass
%and $\sqrt{S}$ the center-of-mass energy of the interaction. This result 
is simple and well behaved, being finite everywhere.

At a deeper thinking, however, problems seem to arise. For instance, one may
ask himself why not to include initial state heavy quarks, coming from the
hadron and to be scattered by the photon, like $\gamma Q \to Q g$. To include
consistently such a diagram is not an easy task, especially if one wants to 
keep the quark massive. Taking it massless, on the other hand, would not only be
a bad approximation but would also produce a divergent total cross section.

\begin{figure}
\begin{center}
\epsfig{file=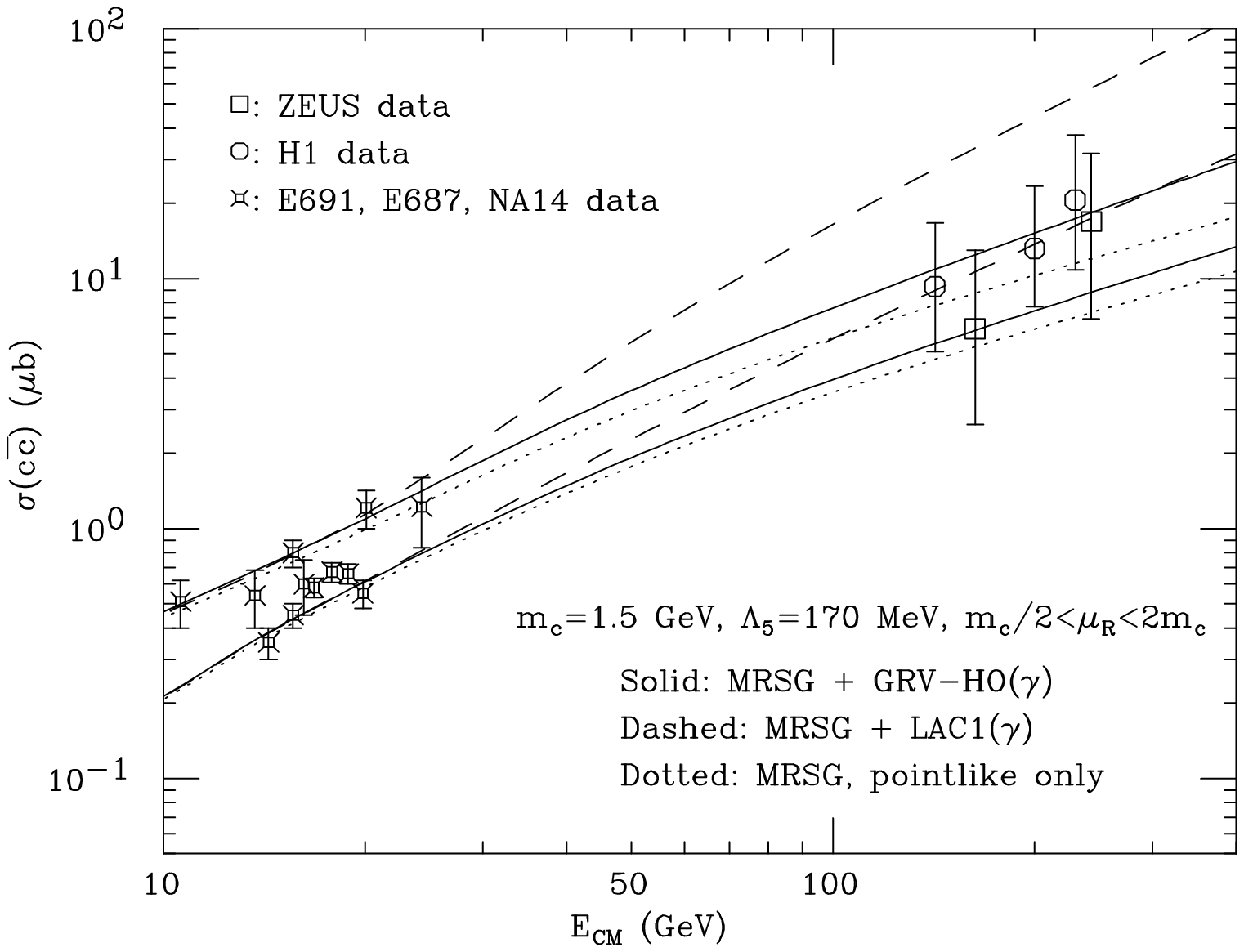,width=10cm,height=6.cm,clip=}
\fcaption{Total cross section for $c\bar c$
photoproduction\protect\cite{fmnr-rev}.}
\label{cctot}
\end{center}
\end{figure}

A way out of this problem was provided by Collins, Soper and Sterman\cite{css}, 
who argued that
the following factorization formula holds for heavy
quarks hadroproduction total cross sections: 
\beq
\sigma(\sqrt{S},m) = \sum_{ij} \int f_{i/H_1} f_{j/H_2} \hat\sigma(ij\to Q\bar
Q;\sqrt{S},m).
\label{QQfact}
\eeq
The sum on the partons runs only on $i$ and $j$ being gluons or light quarks,
and the heavy quarks are only generated at the perturbative level by gluon
splitting. There is therefore no need to try to accommodate them in the
colliding hadrons and the relevant kinematics can be kept exact. 
Eq.~(\ref{QQfact}) provides the basis for an exact perturbative calculation of heavy
quarks production to NLO. For what concerns photoproduction, such a calculation
has been first performed by P.~Nason and K.~Ellis, and subsequently confirmed
by J.~Smith and W.L. van Neerven \cite{en}. 

When going to order $\alpha\as^2$ 
in photon-hadron collision, however, a new feature appears. The photon can now
couple directly to massless quarks, for instance in processes like $\gamma q
\to Q\bar Q q$, and in a given region of phase space a collinear singularity
will appear. It can be consistently factored out, but this requires the
introduction of {\sl photon} parton distribution functions (PDF) 
which, pretty much like the
hadron ones, will describe the probability that before the
interaction the photon splits into hadronic components (light quarks or gluons,
in this case). Such a behaviour is sometimes called {\sl resolved photon} (as
opposed to {\sl direct}). A full NLO calculation for heavy quark
photoproduction will therefore also require a NLO calculation for
hadroproduction\cite{nde}, where one of the PDF's will be the 
photon's one.
A factorization scale $\mu_\gamma$, related to the subtraction of the
singularity at the photon vertex, will link the two pieces and its dependence
on the result will only cancel when both are taken into account.

Frixione, Mangano, Nason and Ridolfi\cite{fmnr} (FMNR) have recently presented 
Montecarlo integrators   for these two calculations, thereby allowing detailed
comparisons with experimental data. A very extensive collection of such
comparisons is presented in a recent review\cite{fmnr-rev}, from which we
select some plots to be shown here.

\begin{figure}
\begin{center}
\epsfig{file=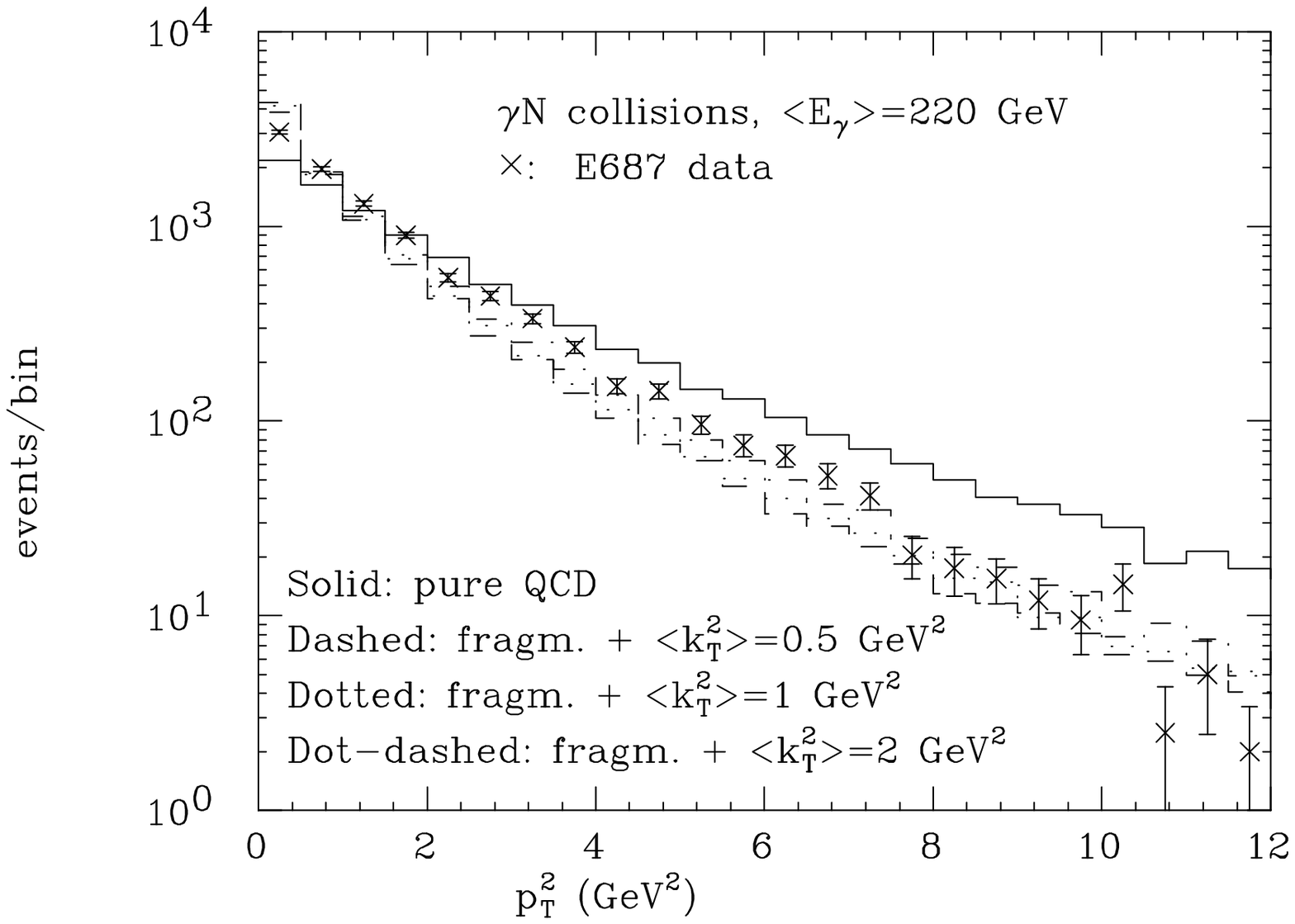,width=7cm,clip=}
\hspace{.5cm}
\epsfig{file=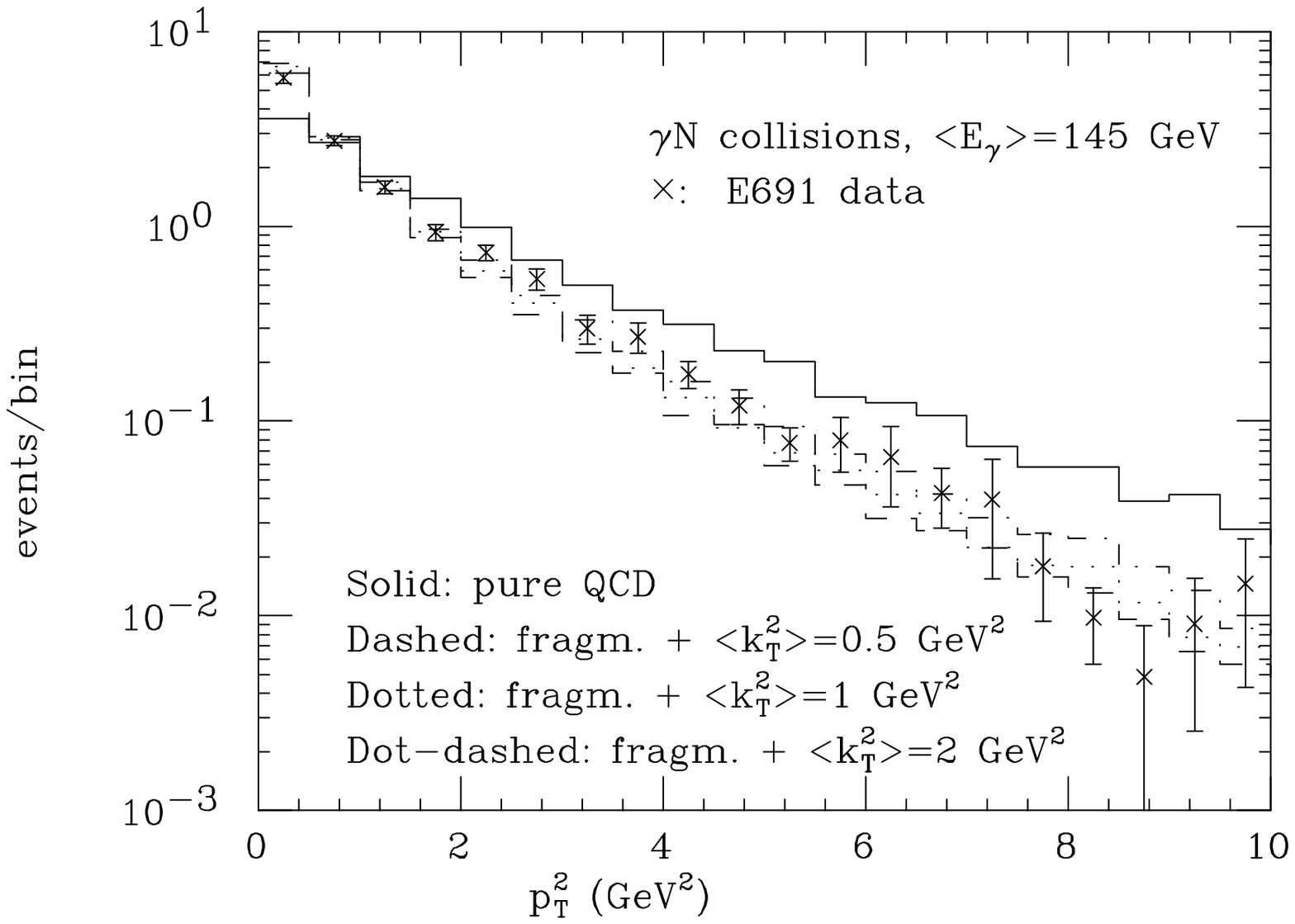,width=7cm,clip=}
\fcaption{Differential $p_T$ distributions for charm production in fixed target
experiments\protect\cite{fmnr-rev}.}
\label{onept}
\end{center}
\end{figure}

A comparison of total cross section experimental results and theoretical
predictions for $c\bar c$ photoproduction is shown in fig. \ref{cctot}.
Although large uncertainties are present, the comparison suggests agreement
between theory and experiment. The new HERA data, at large center of mass
energy, can be seen to appear larger than the pointlike (= direct) photon
prediction only. This suggests the need for a resolved photon component, 
but by no means can determine it precisely.

One-particle transverse momentum ($p_T$) distributions are shown in fig.
\ref{onept}. The pure QCD predictions can be seen to be significantly harder
than the data. However, when corrected with two non-perturbative contributions
they can be matched to the data. These non-perturbative addictions are meant to
represent a primordial transverse momentum $k_T$ of the colliding partons,
other than the one already taken into account by the QCD radiative corrections,
and  the effect of the  fragmentation of the produced heavy quark into the
observed heavy flavoured hadrons, here described by the so-called Peterson
fragmentation function with $\epsilon = 0.06$.

\begin{figure}
\begin{center}
\epsfig{file=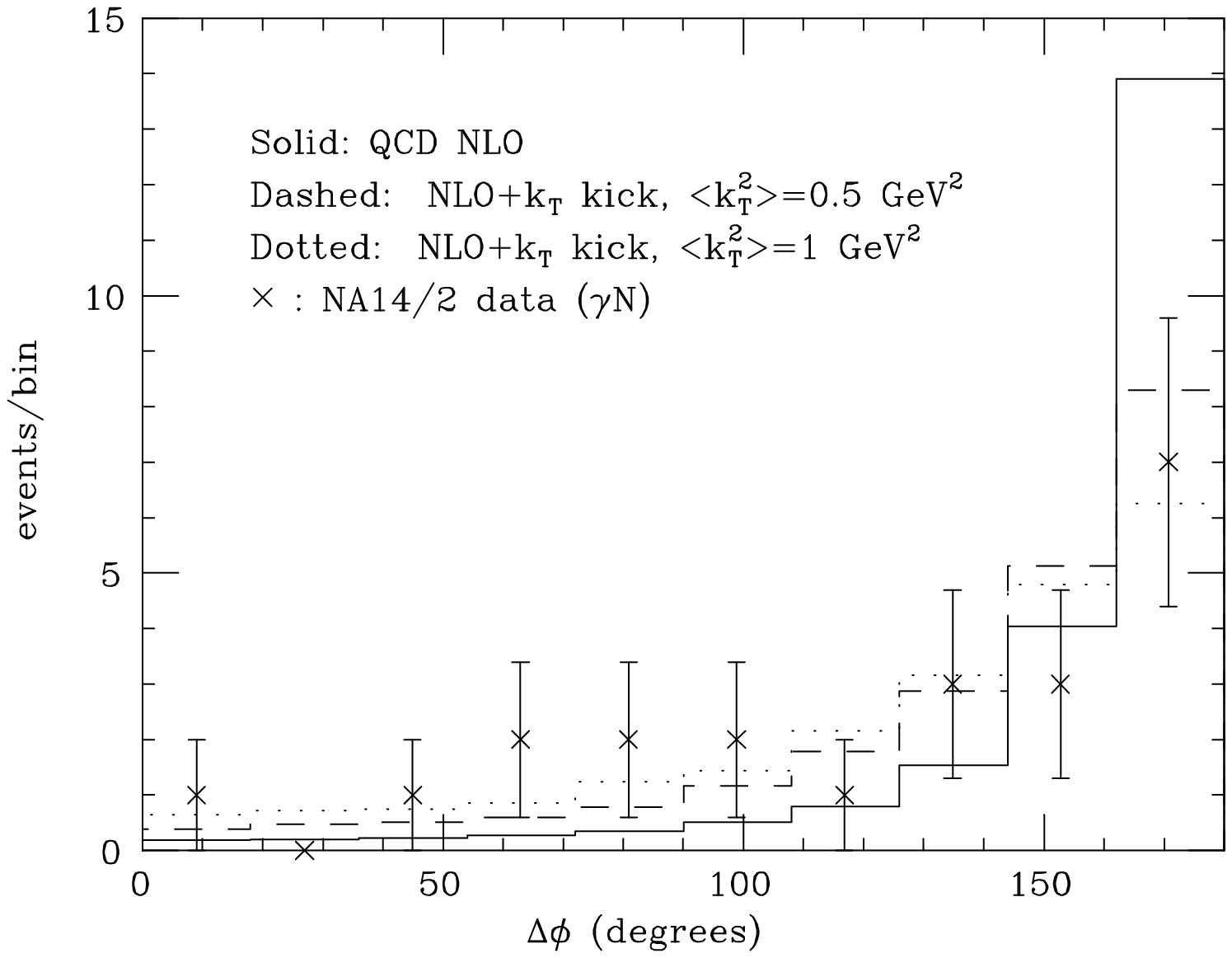,width=7cm,clip=}
\hspace{.5cm}
\epsfig{file=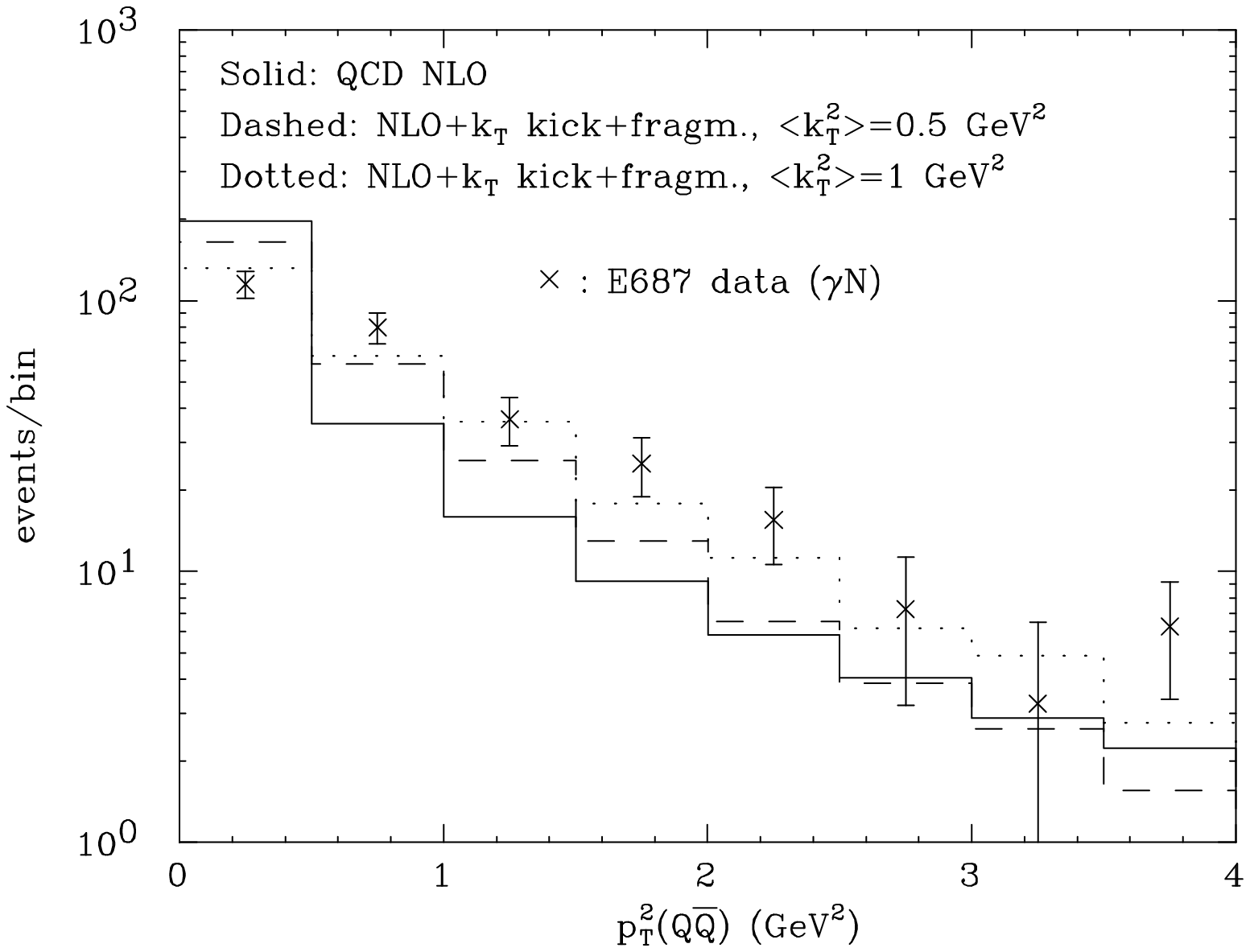,width=7cm,clip=}
\fcaption{Two particles correlations in fixed target 
experiments\protect\cite{fmnr-rev}.}
\label{twopart}
\end{center}
\end{figure}

Comparisons between data and theory for two-particle correlations, like the
azimuthal difference $\Delta\phi$ or the relative transverse momentum
$p_T(Q\overline{Q})$ of the produced heavy quark pair, are shown in fig.
\ref{twopart}. Distributions like these are trivial  in leading order QCD,
since the $Q$ and the $\overline{Q}$ are produced back-to-back. Hence, 
$\Delta\phi = \pi$ and  $p_T(Q\overline{Q}) = 0$. NLO corrections (as well
as non-perturbative contributions) can broaden these distributions, and one
could think of being able to perform a direct measurement of $O(\as^3)$
effects. The plots do however show that non perturbative contributions play a
key role in allowing a good description of the data. One can, however, still
check that the same inputs allow for a good description of both one- and
two-particles distributions, as seems to be the case here.

The overall result of these comparisons  can therefore be summarized as
follows. Total cross sections seem to be well reproduced by the calculation
both at fixed target and HERA regimes, but the huge uncertainties present both
on the experimental and the theoretical side do not allow the study of finer
details like, for instance, the determination of the resolved component at
HERA. For what concerns transverse momentum distributions at fixed target, they
can be reproduced after allowing for  heavy quark fragmentation effects and for
a primordial transverse momentum of the incoming partons of the order of 1 GeV.
These same non-perturbative corrections also allow for a description of
two-particles correlations, thereby pointing towards a consistent picture.

%On the other hand, a word of caution is mandatory in the light of experimental 
%results presented at this Workshop \cite{h1-zeus}, which show a pseudorapidity
%distribution of  the heavy quark at HERA not in agreement with the NLO
%calculation.

\begin{figure}[t]
\begin{center}
\vspace{-1cm}
\epsfig{file=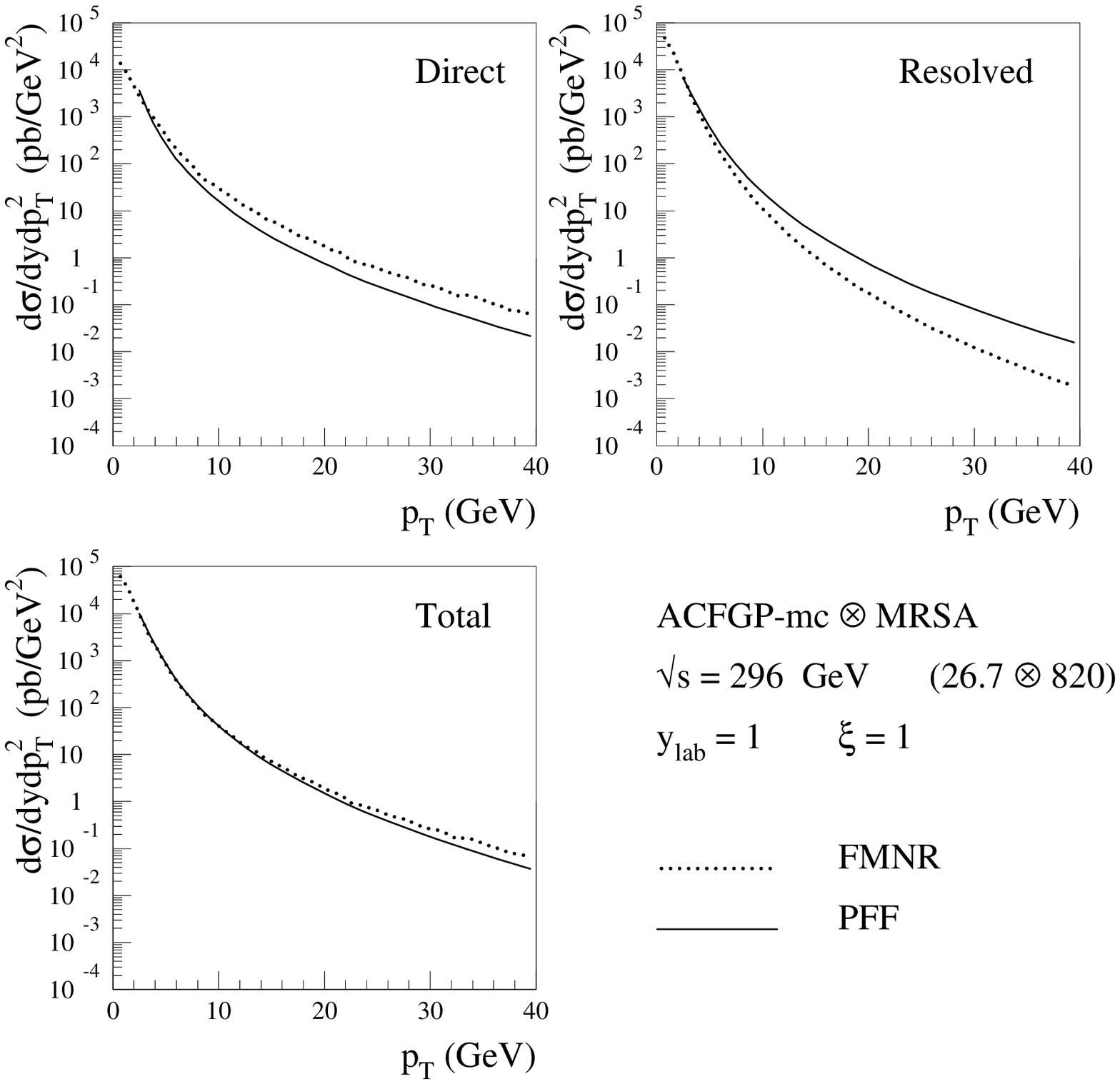,width=13cm,clip=}
\fcaption{Comparison between fixed order (FMNR) and resummed (PFF) 
calculation for charm photoproduction $p_T$ distribution\protect\cite{cg}.
It is worth noticing how the two calculations describe
differently the (unphysical) resolved and direct components, but agree on their
sum (a physical observable).}
\label{ptres}
\end{center}
\end{figure}

\section{Large Transverse Momentum Resummation}
Like any perturbative expansion, the NLO calculation for heavy quarks
photoproduction is only reliable and accurate as long as the coefficients of
the coupling constant remain small. Large terms of the kind $\log(p_T^2/m^2)$
do however appear in the cross section, and for growing $p_T$ they will
eventually became large enough to spoil the convergence of the series. 
Such terms need therefore to be resummed to all orders to allow for a
sensible phenomenological prediction. Such a resummation has been performed
along the following lines\cite{cg}. 

One observes that in the large-$p_T$ limit ($p_T \gg m$) the
only important mass terms are those appearing in the logs, all the others being
power suppressed. This means that an alternative description of heavy quark
production can be achieved by using {\sl massless} quarks and providing at the
same time perturbative distribution and fragmentation functions also for the 
heavy quark, describing the logarithmic mass dependence. The factorization
formula becomes
\beq
d\sigma(p_T) = \sum_{ijk} \int F_{i/H_1}(\mu,[m]) F_{j/H_2}(\mu,[m]) 
d\hat\sigma(ij\to k; p_T, \mu) D_k^Q(\mu,m),
\label{fact}
\eeq
with parton indices $i$,$j$ and $k$ also running on $Q$, taken massless in 
$\hat\sigma$, now an $\overline{\mathrm MS}$ 
subtracted cross section for light partons production.
The dependence on $m$ of the parton distribution functions $F_{i/H}$,  
shown among square brackets in eq. (\ref{fact}), is only there 
when $i$ or $j$ happens to be the heavy quark $Q$.

The key point is that the large mass of the heavy quark allows for 
the evaluation in perturbative QCD (pQCD) of its distribution and fragmentation 
functions.
Initial state conditions for {$F_{{ Q}/H}(\mu_0={ m})$}\cite{ct}
and {$D_k^{ Q}(\mu_0\simeq{ m})$}\cite{melenason}
can be calculated in pQCD at NLO level in the $\overline{\rm MS}$ scheme: 
\begin{eqnarray}
&&F_{Q/H}(x,\mu_0=m) = 0\\
&&D_Q^Q(x,\mu_0) = \delta(1-x) + {{\alpha_s(\mu_0) 
C_F}\over{2\pi}}\left[
{{1+x^2}\over{1-x}}\left(\log{{\mu_0^2}\over{m^2}} -2\log(1-x)
-1\right)\right]_+ \\ 
&&D_g^Q(x,\mu_0) = {{\alpha_s(\mu_0) T_F}\over{2\pi}}
(x^2 + (1-x)^2)
\log{{\mu_0^2}\over{m^2}}  \\
&&D_{q,\bar q,\bar Q}^Q(x,\mu_0) = 0 
\end{eqnarray}
The massive logs will hence appear only through these function, which  can 
then be 
evolved with the Altarelli-Parisi  equations up to the large scale set by $\mu
\simeq p_T$. This evolution will resum to all orders the large logarithms
previously mentioned. 

It is important to mention that due to the neglecting of power suppressed mass
terms this approach becomes unreliable when $p_T\simeq m$. In this region
only a case by case comparison with the full NLO massive calculation -- here
reliable and to be taken as a benchmark -- can tell
how accurate the resummed result is.

Phenomenological analyses show that the effect of the resummation becomes
sizeable only at very large $p_T$, say greater than 20 GeV for charm
photoproduction. Fig.~\ref{ptres} shows the effect of such a resummation for a
fixed photon energy in HERA-like kinematics. The resummed calculation can
be seen to match the fixed order one at $p_T \sim m$, where
resummation effects are not expected to be important, and to behave more softly
in the large $p_T$ region. This particular theoretical refinement should
therefore  not be phenomenologically overly relevant for present-day HERA
physics, data being only available up to $p_T \simeq 12$ GeV.

\section{On the Inclusion of $c\to D^*$ Fragmentation Effects}
When comparing theory with data, one always faces the problem of describing as
closely as possible what the experiments do observe. With heavy quarks
production the problem lies in the experiments actually seeing the decay
products of heavy flavoured hadrons rather than the heavy quark itself. This is
due to the heavy quark strong and non-perturbative binding 
into a hadron prior to decay. This binding involves
the exchange and radiation of low-momentum (order $\Lambda_{QCD}$) gluons, and
typically degrades the momentum of the hadron with respect to the one of the
original quark. Such a degradation can be described with the help of a
non-perturbative fragmentation function (FF) which, lacking the theoretical 
tools to calculate, can be extract by fitting experimental data.

An often employed parametrization for  such a function is the so called
Peterson\cite{pssz} one, which reads
\beq
D_{np}(z;\epsilon) \sim {1\over{z\left[1-1/z-\epsilon/(1-z)\right]^2}}.
\label{peterson}
\eeq
The value of $\epsilon$ is predicted to scale like $\Lambda_{QCD}^2/m^2$. For
charm to $D^*$ fragmentation a global analysis\cite{chrin} based on leading 
order Montecarlo simulations gives the value $\epsilon \simeq 0.06$. This value
has so far usually been taken as the reference one, and used for instance 
together
with the NLO fixed order calculation by FMNR in the plots shown in Section 2.

One should however carefully consider how $\epsilon$ has been extracted from
$e^+e^-$ experimental data.  Experiments usually report the energy or momentum
fraction ($x_E$ or $x_p$) of the observed hadron with respect to the beam
energy. On the other hand the fraction which appears as the argument of the
non-perturbative FF is rather to be taken with respect to the fragmenting quark
momentum, usually denoted by $z$ (see for instance \cite{chrin} for a
discussion on this point). These two fractions are not coincident, due to hard
radiation processes  which lower the momentum of the quark before it fragments
into the hadron. In order to deconvolute these effects one usually runs a
Montecarlo simulation of the collision process at hand, including both the
perturbative parton showers and the subsequent  hadronization of the partons
into the observed hadrons. The latter can be parametrized in the Montecarlo by
the Peterson fragmentation function, and the value of
$\epsilon$ which best describes the data can be extracted. Clearly this
procedure leads to a resulting value for $\epsilon$ which  depends on the
details of the description of the perturbative part. Indeed, the showering
softens the momentum distribution of the heavy quark, producing an effect 
qualitatively similar to that of the non-perturbative FF. On the quantitative
level, the amount of softening (and hence the value of $\epsilon$) required by
the non-perturbative FF to describe  the data is related to the amount of
softening already performed at the  perturbative level. A leading or a
next-to-leading description of the showering can therefore produce different
values for $\epsilon$, whose value is then not a ``unique'' and ``true''one,
but rather closely interconnected  with the details of the description of the
pQCD part of the problem.

\begin{figure}[t]
\begin{center}
\epsfig{file=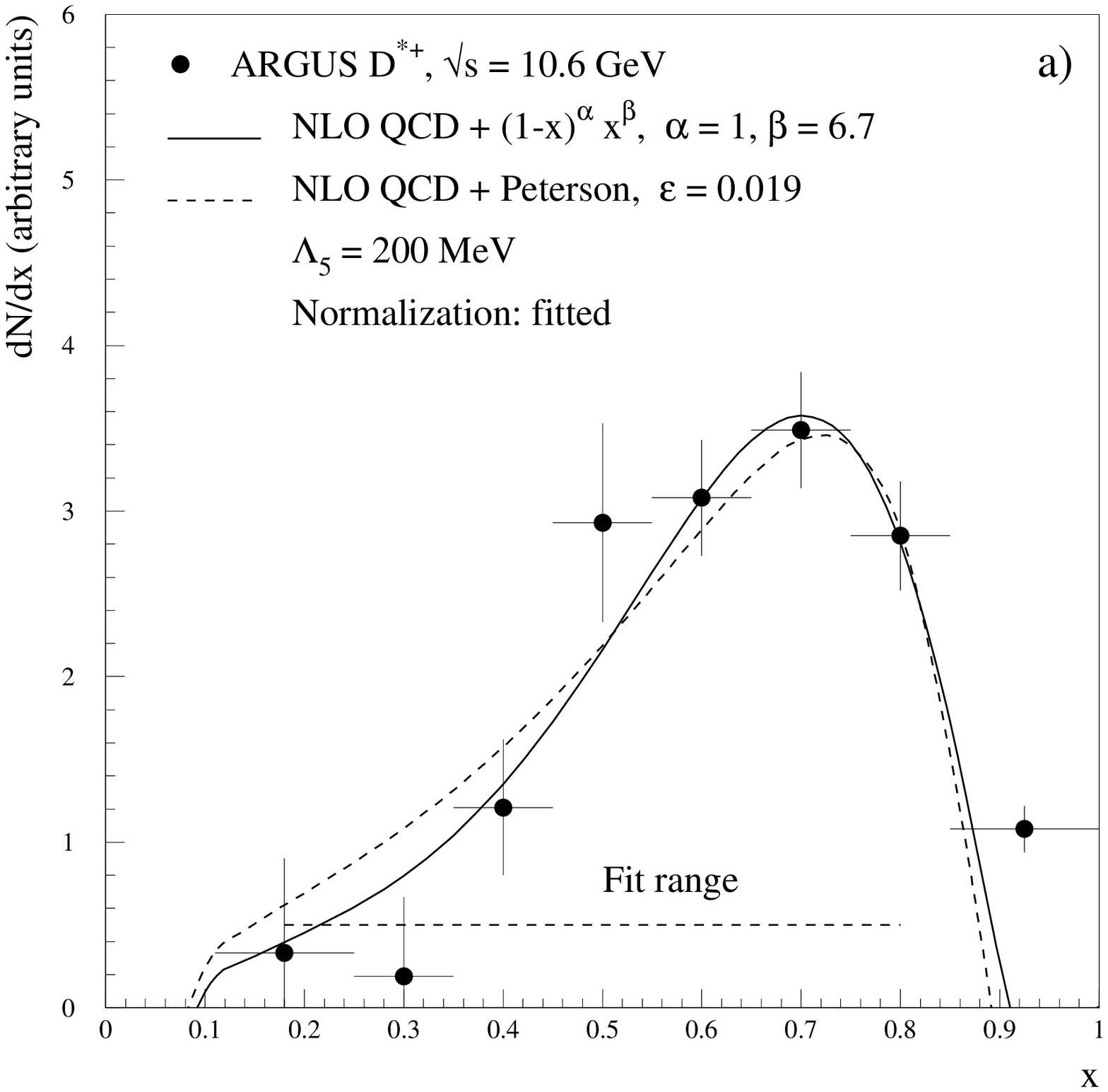,
              bbllx=30pt,bblly=160pt,bburx=540pt,bbury=660pt,
             width=7cm,height=6cm,clip=}
\hspace{.5cm}
\epsfig{file=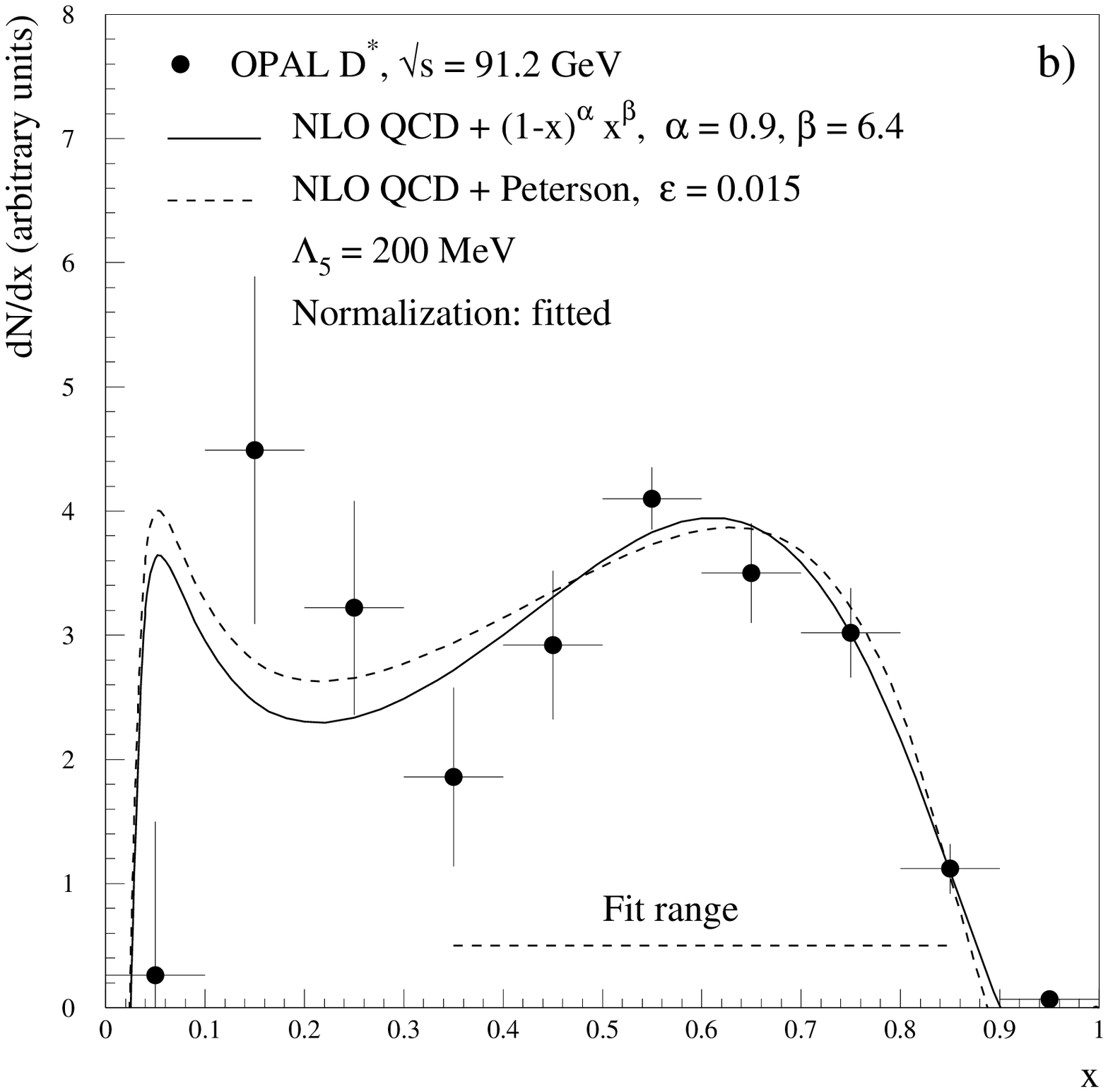,
             bbllx=30pt,bblly=160pt,bburx=540pt,bbury=660pt,
             width=7cm,height=6cm,clip=}
\fcaption{Distributions of $D^*$ mesons as measured by the
ARGUS and OPAL experiments, together with the theoretical
curves\protect\cite{cgee} fitted to 
the same data with
the $(1-x)^\alpha x^\beta$ (full line) and the Peterson (dashed line)
non-perturbative fragmentation functions.}
\label{argusopal}
\end{center}
\end{figure}

In ref. \cite{cgee} fits to $D^*$ data taken by the ARGUS and OPAL experiments
have been performed with NLO accuracy using a fragmentation description for 
the heavy quark
production like the one described in Section 3, complemented with the 
inclusion of a non-perturbative component via the ansatz
\beq
D_k^{D^*}(\mu) = D_k^c(\mu) \otimes D_c^{D^*},
\label{ansatz}
\eeq
represented by the convolution of a perturbatively calculable fragmentation
function of the parton $k$ into the heavy quark $c$ and the non-perturbative
form $D_c^{D^*}$ describing the $c\to D^*$ transition. This non perturbative
form is taken to be scale independent, i.e. all scaling effects are assumed to
be described by the Altarelli-Parisi evolution of the perturbative part
$D_k^c(\mu)$. A similar approach had already been introduced in \cite{melenason}.

Results for these fits are shown in fig. \ref{argusopal}. The value for
$\epsilon$ has been consistently found to be of order 0.02 rather than the
customary 0.06 one, resulting instead from fits with leading order evolution. 
Recalling the previous discussion, this comes to no surprise: next-to-leading
order evolution softens more the heavy quark spectrum, and a harder
non-perturbative fragmentation function is therefore needed to provide a
satisfactory description of the data (see \cite{cgee} for a full discussion).

Similar fits to $e^+e^-$ data have also been performed by Binnewies, Kniehl and
Kramer\cite{bkk} (BKK). These authors do instead find, again with NLO
evolution, a value for $\epsilon$ still 
close to the usual 0.06. This discrepancy, beyond irrelevant nomenclature
differences,  can be traced back to a discrepancy in the implementation of
the factorization scheme. The scheme used in \cite{cgee}, as originally set up
in \cite{melenason}, is the customary $\overline{\rm MS}$ one. Considering for
instance the
dominant non-singlet component only for simplicity, the $e^+e^-\to QX$ momentum
distribution $d\sigma/dx$ can be schematically written as the convolution 
(= product in Mellin
moments space) of a short distance coefficient function, an Altarelli-Parisi
evolution kernel $E(\mu,\mu_0)$, a perturbative initial state condition for 
the heavy quark perturbative
fragmentation function (PFF) and a fixed non-perturbative FF,
\beq
d\sigma(\sqrt{S},m) = \Big(1 + \alpha_s(\mu) c(\sqrt{S},\mu)\Big) 
         E(\mu,\mu_0) \Big(1+\alpha_s(\mu_0) d(\mu_0,m)\Big) D_{np},
\label{cgmn}
\eeq
where the perturbative expansions of the coefficient function and the PFF have
been explicitly shown. The factorization scale $\mu$ is taken of the order of
the (large) collision energy $\sqrt{S}$, and the initial scale $\mu_0$ is 
taken of the order of the quark mass $m$.

BKK on the other hand, employing a scheme introduced by Kniehl, Kramer and
Spira\cite{kks} (KKS), write $d\sigma(\sqrt{S},m)$ as
\beq
d\sigma(\sqrt{S},m) = \Big(1 + \alpha_s(\mu) c(\sqrt{S},\mu) + \alpha_s(\mu)
        d(\mu_0,m)\Big) E(\mu,\mu_0) D_{np}.
\label{bkks}
\eeq
These two expressions can be seen to differ by $O(\as^2)$ terms. However, one
of these terms is given by
\beq
\alpha_s(\mu) - \alpha_s(\mu_0) = -b_0 \alpha_s^2 \log\frac{\mu^2}{\mu_0^2}
\eeq
and is, therefore, one of the next-to-leading logarithms (NLL) 
$\alpha_s^k \log^{k-1}(\sqrt{S}/m)$ we are resumming. Hence the two calculations
differ by a NLL term and cannot possibly {\sl both} implement
correctly a resummation at the NLL level. 

To better understand the discrepancy, the BKKS scheme can for instance be
rewritten in the form (\ref{cgmn}), with an initial state condition for the
PFF containing the large scale $\mu$ as the argument for $\alpha_s$ rather
than the small one $\mu_0$. This choice of a large scale is however in
contradiction with the  factorization theorem hypotheses, which only allow for
small scales in initial conditions, to avoid the appearance of unresummed
large logs. Choosing the large $\mu$ leads at a practical level to  the
difference being reabsorbed into a  different value  for the $\epsilon$
parameter, which happens quite accidentally to be 0.06  rather than 0.02.  One
can show that, replacing in the BKKS formula (\ref{bkks}) the $\alpha_s(\mu)
d(\mu_0,m)$ term with  $\alpha_s(\mu_0) d(\mu_0,m)$ (or alternatively
appropriately 
modifying the NLO splitting vertices in the evolution kernel),  $\epsilon =
0.02$ is once again found from the NLL fits within this scheme too.

\begin{figure}[t]
\begin{center}
\epsfig{file=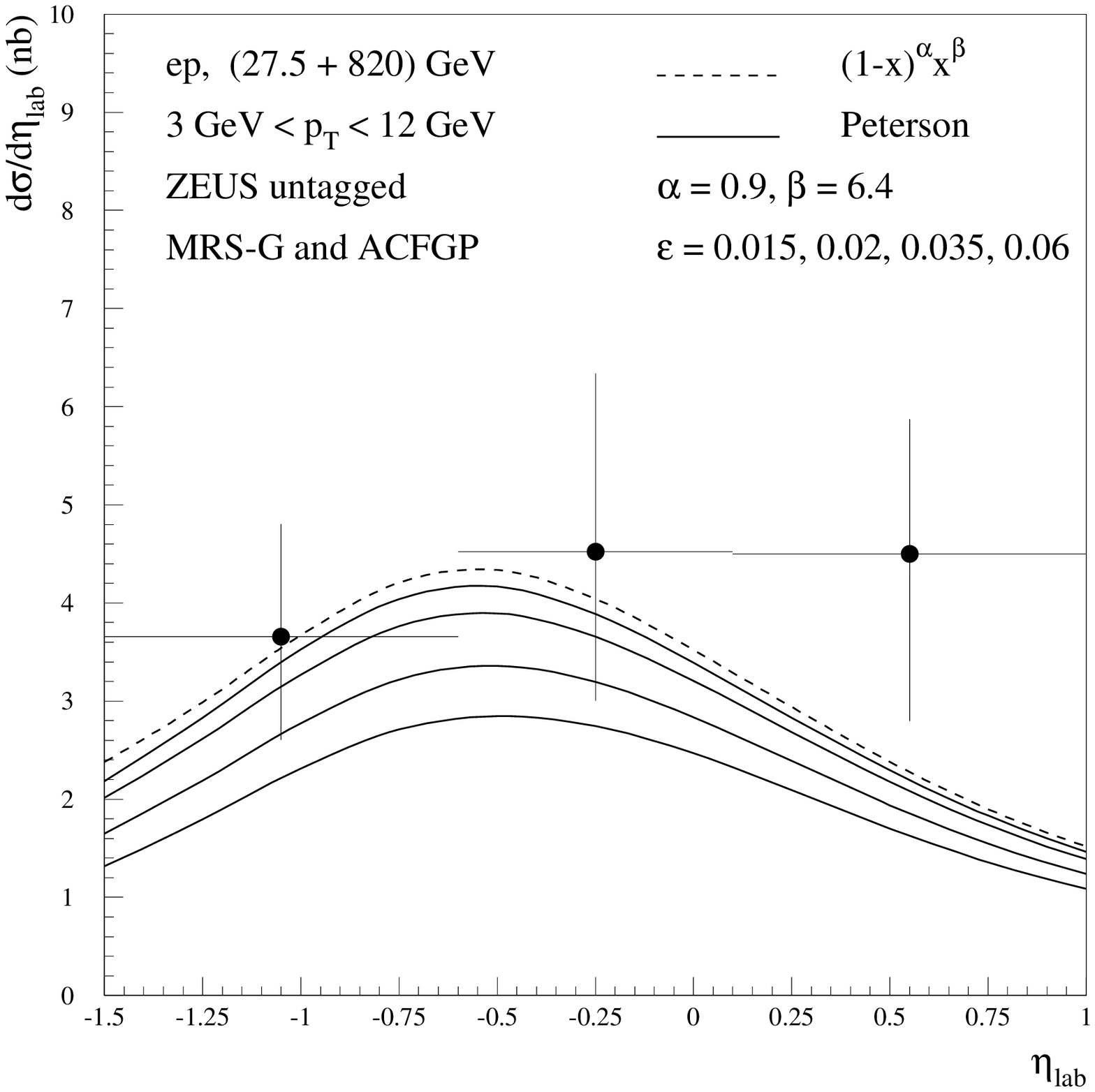,
              bbllx=30pt,bblly=160pt,bburx=540pt,bbury=660pt,
             width=7cm,height=6cm,clip=}
\hspace{.5cm}
\epsfig{file=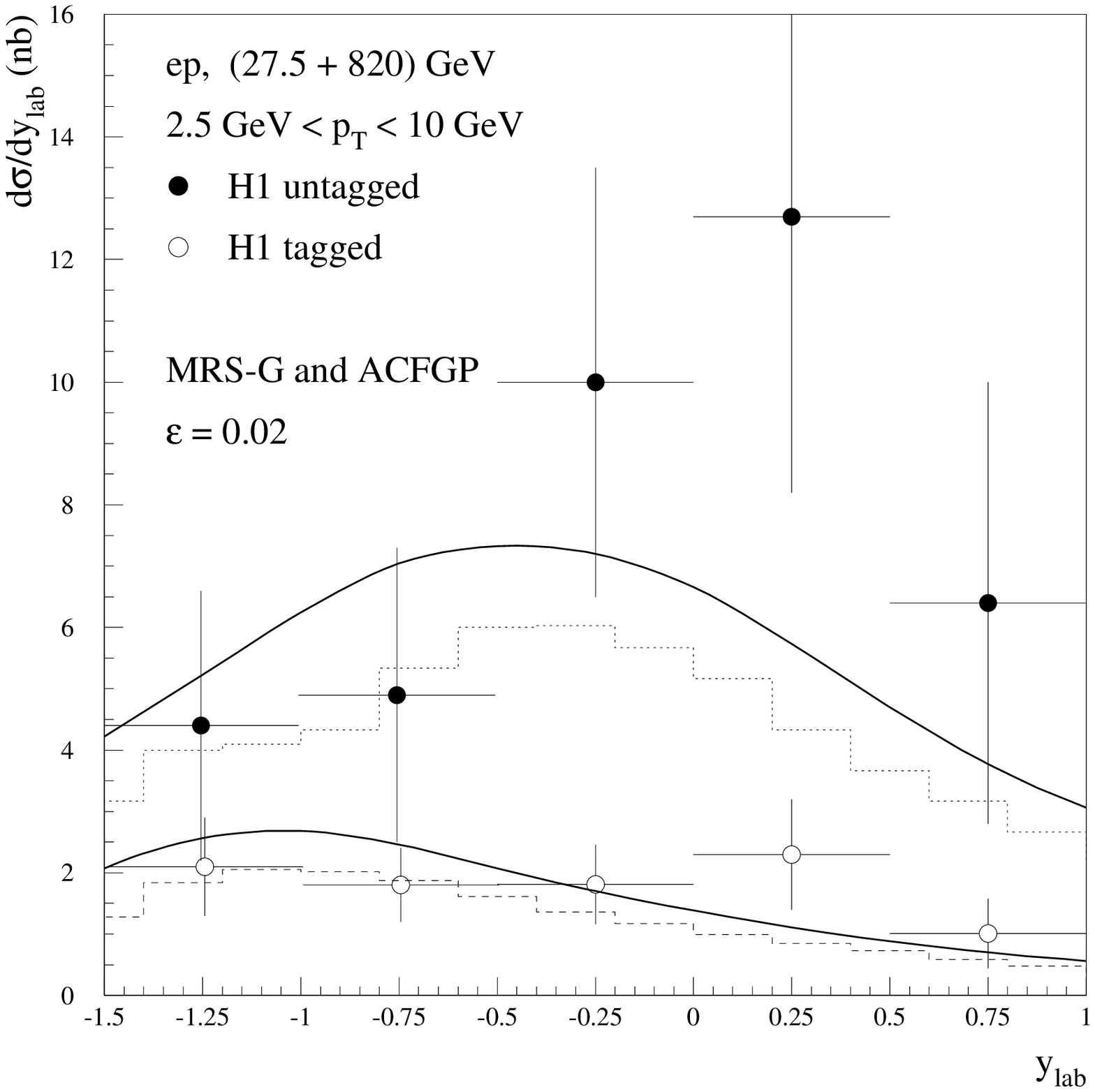,
             bbllx=30pt,bblly=160pt,bburx=540pt,bbury=660pt,
             width=7cm,height=6cm,clip=}
\fcaption{Effect on the $D^*$ photoproduction cross section of a decreasing
value for $\epsilon$ (left), and comparison of H1 data with the fixed order
calculation by FMNR (histograms, $\epsilon = 0.06$) and the fragmentation 
function approach (smooth lines, $\epsilon = 0.02$).}
\label{h1}
\end{center}
\end{figure}

On the phenomenological side, and making use of the universality argument, one
can now argue that the use of a ``harder'' Peterson form with $\epsilon=0.02$
is probably more suited when combined with a NLO perturbative calculation like
the FMNR one which, albeit only at fixed order, contains NLL gluon radiation.
Decreasing $\epsilon$ means increasing the cross section at large $p_T$, being
the $p_T$ distribution steeply falling with increasing transverse momentum.
This could help reconciling the HERA experimental data\cite{h1-zeus}  with the
perturbative NLO calculation, which was shown to underestimate them a little
when convoluted with a Peterson with $\epsilon=0.06$: fig. \ref{h1} shows, on
the left, how the cross section for $D^*$ photoproduction at HERA increases
with decreasing $\epsilon$ and, on the right, a comparison of the H1 data
with the fixed order prediction  by FMNR ($\epsilon = 0.06$) and the
fragmentation functions one with $\epsilon = 0.02$. One should notice that the
$p_T$ values involved are still pretty small: this means that the fixed order
calculation is still reliable and the accuracy of the resummed one has to be
assessed first by comparing with the former. In this case they are found to be
in good agreement,  the difference in the plot being mainly given by the
different $\epsilon$  values.

Last but not least, it is worth mentioning how, going from LO to NLO analyses, 
a similar hardening of the non-perturbative
fragmentation function is also expected for the
$b$ quark. The corresponding increase of the hadroproduction bottom $p_T$
distributions\cite{mlm} 
would be welcome in the light of the Tevatron data presently overshooting the
theoretical predictions by at least 30\%.

{
\vspace{.4cm}\noindent
{\bf Acknowledgements.} I wish to thank the Organizers of this Workshop for
the invitation to give this talk,
Mario Greco for his collaboration and Paolo Nason for the many conversations 
on the heavy quarks physics items I've been reviewing here.
}

%\vspace{-.2cm}
\section{References}
% ----------------------------------------------------------------------------
% ------------------- define commands for some papers -------------------
\newcommand{\zp}[3]{{\it Zeit.\ Phys.\ }{\bf C#1} (19#2) #3}
\newcommand{\pl}[3]{{\it Phys.\ Lett.\ }{\bf B#1} (19#2) #3}
\newcommand{\plold}[3]{{\it Phys.\ Lett.\ }{\bf #1B} (19#2) #3}
\newcommand{\np}[3]{{\it Nucl.\ Phys.\ }{\bf B#1} (19#2) #3}
\newcommand{\prd}[3]{{\it Phys.\ Rev.\ }{\bf D#1} (19#2) #3}
\newcommand{\prl}[3]{{\it Phys.\ Rev.\ Lett.\ }{\bf #1} (19#2) #3}
\newcommand{\prep}[3]{{\it Phys.\ Rep.\ }{\bf C#1} (19#2) #3}
\newcommand{\niam}[3]{{\it Nucl.\ Instr.\ and Meth.\ }{\bf #1} (19#2) #3}
\newcommand{\mpl}[3]{{\it Mod.\ Phys.\ Lett.\ }{\bf A#1} (19#2) #3}

\end{document}